\documentclass[onecolumn,showpacs,preprintnumbers]{revtex4}
\usepackage{graphicx}
\usepackage{dcolumn}
\usepackage{bm}
\usepackage{epsfig}
\usepackage{amsmath,amssymb}

\begin{document}
\title{Analysis of the strong coupling form factors of $\Sigma_bNB$ and $\Sigma_c ND$ in QCD sum rules}
\author{Guo-Liang Yu$^{1}$}
\email{yuguoliang2011@163.com}
\author{Zhi-Gang Wang$^{1}$}
\email{zgwang@aliyun.com}
\author{Zhen-Yu Li$^{2}$}

\affiliation{$^1$ Department of Mathematics and Physics, North China
Electric power university, Baoding 071003, People's Republic of China\\$^2$
School of Physics and Electronic Science, Guizhou Normal College,
Guiyang 550018, People's Republic of China}
\date{\today }

\begin{abstract}
In this article, we study the strong interaction of the vertexes
$\Sigma_bNB$ and $\Sigma_c ND$ using the three-point QCD sum rules
under two different dirac structures. Considering the contributions
of the vacuum condensates up to dimension $5$ in the operation
product expansion, the form factors of these vertexes are
calculated. Then, we fit the form factors into analytical functions
and extrapolate them into time-like regions, which giving the
coupling constant. Our analysis indicates that the coupling constant
for these two vertexes are $G_{\Sigma_bNB}=0.43\pm0.01GeV^{-1}$ and
$G_{\Sigma_cND}=3.76\pm0.05GeV^{-1}$.
\end{abstract}


\maketitle

\begin{large}
\textbf{1 Introduction}
\end{large}

By this time, many heavy baryons have been observed by BaBar, Belle
and CLEO Collaborations\cite{Aube06,Naka10,Lesi,Rosn07} such as
$\frac{1}{2}^{+}$ antitriplet
states($\Lambda^{+}_{c}$,$\Xi^{+}_{c}$,$\Xi^{0}_{c}$), the
$\frac{1}{2}^{+}$ and $\frac{3}{2}^{+}$ sextet states
($\Omega_{c}$,$\Sigma_{c}$,$\Sigma'_{c}$) and
($\Omega^{*}_{c}$,$\Sigma^{*}_{c}$,$\Sigma^{*}_{c}$)\cite{Naka10}.
Besides, several S-wave bottom baryon states such as
$\Lambda_{b}$,$\Sigma_{b}$,$\Sigma^{*}_{b}$,$\Xi_{b}$ and
$\Omega_{b}$ have also been observed by CDF and LHCb
Collaborations\cite{Paul,Klem10}. The SELEX collaboration even have
reported the observation of the signal for the doubly charmed baryon
state $\Xi_{cc}^{+}$\cite{Matt,Oche}. Since then, people have showed
great interest in studying the properties of these heavy baryons
which contains at least a heavy
quark\cite{Faes,Pate,LiuX,Chun,Zhjr}. The charm and bottom baryon
states which contain one (two) heavy quark(s) are particularly
interesting for studying dynamics of the light quarks in the
presence of the heavy quark(s), and serve as an excellent ground for
testing predictions of the quark models and heavy quark symmetry.

The properties of the heavy baryons such as mass spectrum, radiative
and strong decays have been studied by many researches, which is
very important for us to further understand the heavy flavor
physics\cite{Karl1,Karl2,Nava,Khod,Alie,Azi1,Azi2,Wzg}. In this
regards, the strong coupling constants associating with heavy
baryons play an important role in describing the strong interaction
among the heavy baryons and other participated hadrons. In addition,
the properties of $B$ and $D$ mesons in nuclear medium are closely
related with their interactions with the
nucleons\cite{WzgH,Kuma,Haya}, i.e.

 $D^{0}+p$  or    $n\rightarrow\Lambda_{c}^{+},\Sigma_{c}^{+}$ or
$\Sigma_{c}^{0}$

$ B^{-}+p$  or   $n\rightarrow\Lambda_{b}^{0}$ or $\Sigma_{b}^{-}$

From these processes, we can see that it is significant to know the
values of the related strong coupling constants $G_{\Sigma_bNB}$ and
$G_{\Sigma_c ND}$ which is essential to determine the modifications
on the masses, decay constants and other parameters of the $B$ and
$D$ mesons in nuclear medium. Up to now, only a few works on the
strong coupling constants of the heavy baryons with the nucleon and
heavy mesons have been reported\cite{Azi1,Azi2,Navar,Khodj}.

On the other hand, QCD sum rules is one of the most powerful
non-perturbative methods, which is also independent of model
parameters. In recent years, numerous research articles have been
reported about the precise determination of the strong form factors
and coupling constants via
QCDSR\cite{Brac1,Brac2,Wzg2,Khod2,Khod3,Alie2,Alie3,Doi,Altm,Wzg3,Cerq,Rodr,Yazi,Khos1,Khos2,Rein,Pasc,Wzg4}.
In this work, we use the QCDSR formalism to obtain the coupling
constants of the strong vertexs $\Sigma_bNB$ and $\Sigma_c ND$,
where the contributions of vacuum condensates up to 5 in the OPE are
considered.

The outline of this paper is as follows. In Sect.2, we study the
strong vertexs $\Sigma_bNB$ and $\Sigma_c ND$ using the three-point
QCDSR under two different dirac structures $p\!\!/\gamma_{5}$ and
$q\!\!/\gamma_{5}$. Besides of the perturbative contribution, we
also consider the contributions of $\langle\overline{q}q\rangle$,
$\langle\overline{G}G\rangle$ and $\langle qqG\rangle$ at OPE side.
In Sect.3, we present the numerical results and discussions, and
Sect.4 is reserved for our conclusions.

\begin{large}
\textbf{2 QCD sum rules for $\Sigma_bNB$ and $\Sigma_c ND$}
\end{large}

The three-point correlation functions of these two vertices
$\Sigma_bNB$ and $\Sigma_c ND$ can be written as:
\begin{eqnarray}
\Pi(p,p',q)=i^{2}\int d^{4}x\int
d^{4}ye^{-ip.x}e^{ip'.y}\Big\langle0|\tau(J_{N}(y)J_{B[D]}(0)\overline{J}_{\Sigma_{b}[\Sigma_{c}]}(x))
|0\Big\rangle
\end{eqnarray}

Where $\tau$ is the time ordered product and
$J_{\Sigma_{b}[\Sigma_{c}]}(x)$,$J_{N}(y)$ and $J_{B[D]}(0)$ are the
interpolating currents of the hadrons $\Sigma_{b}[\Sigma_{c}]$, $N$
and $B[D]$ respectively:
\begin{eqnarray}
&& J_{\Sigma_{b}[\sum_{c}]}(x)=\epsilon_{ijk}\Big(u^{iT}(x)C\gamma_{\mu}d^{j}(x)\Big)\gamma_{5}\gamma^{\mu}b[c]^{k}(x)\\
&& J_{N}(y)=\epsilon_{ijk}\Big(u^{iT}(y)C\gamma_{\mu}u^{j}(y)\Big)\gamma_{5}\gamma^{\mu}d^{k}(y)\\
&& J_{B[D]}(0)=\overline{u}(0)\gamma_{5}b[c](0)
\end{eqnarray}
where $C$ is the charge conjugation operator, and $i$, $j$ and $k$
are color indices.

According to the QCD sum rules, the three-point correlation function
can be calculated in two different ways. In the first way, the
calculation is carried out in hadron degrees of freedom, called the
phenomenological side. Secondly, it is called OPE side which is
calculated in quark degrees of freedom. Then, invoking the
quark-hadron duality, we equate the phenomenological and OPE sides
from which the QCD sum rules for the strong coupling form factors is
attained.

\begin{large}
\textbf{2.1 The phenomenological side}
\end{large}

We insert a complete set of intermediate hadronic states with the
same quantum numbers as the operators
$J_{\Sigma_{b}[\Sigma_{c}]}(x)$,$J_{N}(y)$ and $J_{B[D]}(0)$ into
the correlation function Eq$(1)$ to obtain the phenomenological
representations. After isolating the ground-state contributions, the
correlation function is written as:
\begin{eqnarray}
\notag
 \Pi^{HAD}(p,p',q)=&&\frac{\Big \langle 0| J_{N}|N(p')\Big
\rangle \Big \langle 0| J_{B[D]}|B[D](q)\Big \rangle \Big \langle
\Sigma_{b}[\Sigma_{c}](p)|
\overline{J}_{\Sigma_{b}[\Sigma_{c}]}|0\Big
\rangle}{(p^{2}-m^{2}_{\Sigma_{b}[\Sigma_{c}]})(p'^{2}-m^{2}_{N})(q^{2}-m^{2}_{B[D]})}\\
&& \Big\langle
N(p')B[D](q)|\Sigma_{b}[\Sigma_{c}](p)\Big\rangle+\cdots
\end{eqnarray}
Where $h.r.$ stands for the contributions of higher resonances and
continuum states. And the matrix elements appearing in the above
equation can be parameterized as the following formulas:
\begin{eqnarray}
\langle 0| J_{N}|N(p') \rangle=&&\lambda_{N}u_{N}(p',s')\\
  \langle
0|J_{B[D]}|B[D](q) \rangle=&&i\frac{m_{B[D]}^{2}f_{B[D]}}{m_{u}+m_{b[c]}}\\
 \langle \Sigma_{b}[\Sigma_{c}](p)|
\overline{J}_{\Sigma_{b}[\Sigma_{c}]}|0
\rangle=&&\lambda_{\Sigma_{b}[\Sigma_{c}]}\overline{u}_{\Sigma_{b}[\Sigma_{c}]}(p,s)\\
\langle
N(p')B[D](q)|\Sigma_{b}[\Sigma_{c}](p)\rangle=&&G_{\Sigma_{b}NB[\Sigma_{c}ND]}\overline{u}_{N}(p',s')i\gamma_{5}u_{\Sigma_{b}[\Sigma_{c}]}(p,s)
\end{eqnarray}
Where $\lambda_{N}$ and $\lambda_{\Sigma_{b}[\Sigma_{b}]}$ are
residues of $N$ and $\Sigma_{b}[\Sigma_{b}]$ baryons, $f_{B[D]}$ is
the leptonic decay constant of $B[D]$ meson and
$G_{\Sigma_{b}NB[\Sigma_{c}ND]}$ is the strong coupling form factor
of the vertices $\Sigma_bNB$ and $\Sigma_c ND$. Considering these
parameters, Eq.$(5)$ can be written as:
\begin{eqnarray}
\notag
\Pi^{HAD}(p,p',q)=&&i^{2}\frac{m_{B[D]}^{2}f_{B[D]}}{m_{b[c]}+m_{u}}\frac{\lambda_{N}\lambda_{\Sigma_{b}[\Sigma_{c}]}g_{\Sigma_{b}NB[\Sigma_{c}ND]}}{(p^{2}-m^{2}_{\Sigma_{b}[\Sigma_{c}]})(p'^{2}-m^{2}_{N})(q^{2}-m^{2}_{B[D]})}\\
&& \notag \times\Big\{
(m_{N}m_{\Sigma_{b}[\Sigma_{c}]}-m_{\Sigma_{b}[\Sigma_{c}]}^{2})\gamma_{5}+(m_{\Sigma_{b}[\Sigma_{c}]}-m_{N})p\!\!\!/\gamma_{5}+q\!\!\!/p\!\!\!/\gamma_{5}-m_{\Sigma_{b}[\Sigma_{c}]}q\!\!\!/\gamma_{5}\Big\}\\
&&+\cdots
\end{eqnarray}

\begin{large}
\textbf{2.2 The OPE side}
\end{large}

Now, we briefly outline the operator product expansion(OPE) for the
three-point correlation Eq.$(1)$. Firstly, we contract the quark
fields in the correlation with Wich's theorem.
\begin{eqnarray}
\notag\ \Pi(p,p',q)^{OPE}=&&i^{2}\int d^{4}x\int
d^{4}ye^{-ip.x}e^{ip'.y}\epsilon_{abc}\epsilon_{ijk}\\
&& \notag
\times\Big\{\gamma_{5}\gamma_{\nu}S_{d}^{cj}(y-x)\gamma_{\mu}CS_{u}^{biT}(y-x)C\gamma_{\nu}S_{u}^{ah}(y)\gamma_{5}S_{b[c]}^{hk}(-x)\gamma_{\mu}\gamma_{5}\\
&&-\gamma_{5}\gamma_{\nu}S_{d}^{cj}(y-x)\gamma_{\mu}CS_{u}^{aiT}(y-x)C\gamma_{\nu}S_{u}^{bh}(y)\gamma_{5}S_{b[c]}^{hk}(-x)\gamma_{\mu}\gamma_{5}\Big\}
\end{eqnarray}
Secondly, we replace the heavy and light quark propagators with the
following full propagators\cite{Rein,Pasc,ZGW14},
\begin{eqnarray}
\notag
 S_{b[c]}^{mn}(x)=&&\frac{i}{(2\pi)^{4}}\int
 d^{4}ke^{-ik.x}\Big\{\frac{\delta_{mn}}{k\!\!\!/-m_{b[c]}}-\frac{g_{s}G_{mn}^{\alpha\beta}}{4}\frac{\sigma_{\alpha\beta}(k\!\!\!/+m_{b[c]})+(k\!\!\!/+m_{b[c]})\sigma_{\alpha\beta}}{(k^{2}-m_{Q}^{2})^{2}}\\
 &&+\frac{\pi^{2}}{3}\Big\langle\frac{\alpha_{s}GG}{\pi}\Big\rangle\delta_{mn}m_{b[c]}\frac{k^{2}+m_{b[c]}k\!\!\!/}{(k^{2}-m_{b[c]}^{2})^{4}}+\cdots\Big\}
\end{eqnarray}
\begin{eqnarray}
 S_{u[d]}^{mn}(x)=&&i\frac{x\!\!\!/}{2\pi^{2}x^{4}}\delta_{mn}-\frac{m_{u[d]}}{4\pi^{2}x^{2}}\delta_{mn}-\frac{\langle
 \overline{q}q\rangle}{12}\Big(1-i\frac{m_{u[d]}}{4}x\!\!\!/\Big)-\frac{x^{2}}{192}m_{0}^{2}\langle
 \overline{q}q\rangle\Big( 1-i\frac{m_{u[d]}}{6}x\!\!\!/\Big)\\
 &&
 \notag-\frac{ig_{s}\lambda_{A}^{ij}G^{A}_{\theta\eta}}{32\pi^{2}x^{2}}\Big[x\!\!\!/\sigma^{\theta\eta}+\sigma^{\theta\eta}x\!\!\!/\Big]+\cdots
\end{eqnarray}
Where $m$,$n$ are the color indices, and $\langle
q\overline{q}\rangle$ is the $\langle u\overline{u}\rangle$ and
$\langle d\overline{d}\rangle$ in Eq$(13)$. After these above
substitutions in Eq(11), we carry out Fourier transformation in D=4
dimensions using the following formulas:
\begin{eqnarray}
\frac{1}{[(y-x)^{2}]^{n}}=\int\frac{d^{D}t}{(2\pi)^{D}}e^{-it.(y-x)}i(-1)^{n+1}2^{D-2n}\pi^{D/2}\frac{\Gamma(D/2-n)}{\Gamma(n)}\Big(-\frac{1}{t^{2}}\Big)^{D/2-n}
\end{eqnarray}
\begin{eqnarray}
\frac{1}{[y^{2}]^{n}}=\int\frac{d^{D}t'}{(2\pi)^{D}}e^{-it'.y}i(-1)^{n+1}2^{D-2n}\pi^{D/2}\frac{\Gamma(D/2-n)}{\Gamma(n)}\Big(-\frac{1}{t'^{2}}\Big)^{D/2-n}
\end{eqnarray}
Before the  preformation of four-$x$ and four-$y$ integrals, the
replacements $x_{\mu}$ $\rightarrow$ $i\frac{\partial}{\partial
p_{\mu}}$ and $y_{\mu}$ $\rightarrow$ $-i\frac{\partial}{\partial
p'_{\mu}}$ are carried out. After these processes, the integrals
turn into Dirac delta functions which are used to simplify the
four-integrals over $k$ and $t'$. The following step is to perform
the Feynman parametrization, after which the following function is
used to carried out the remaining four-integral over $t$.
\begin{eqnarray}
\notag\
\int\frac{d^{D}t}{(2\pi)^{D}}\frac{1}{[t-M^{2}]^{\alpha}}&&=\frac{i(-1)^{\alpha}}{(4\pi)^{D/2}}\frac{\Gamma(\alpha-D/2)}{\Gamma(\alpha)}\frac{1}{(M^{2})^{\alpha-D/2}}
\end{eqnarray}
Where $M^{2}=m_{b[c]}^{2}x+p^{2}x(x+y-1)+p'^{2}y(x+y-1)-q^{2}xy$.

After further simplification, the three-point correlation in OPE
side show the following Dirac structures:
\begin{eqnarray}
\Pi^{OPE}(p,p',q)=\Pi_{1}(q^{2})\gamma_{5}+\Pi_{2}(q^{2})p\!\!\!/\gamma_{5}+\Pi_{3}(q^{2})q\!\!\!/p\!\!\!/\gamma_{5}+\Pi_{4}(q^{2})q\!\!\!/\gamma_{5}
\end{eqnarray}
Where each $\Pi_{i}$ denotes contributions coming from perturbative
and nonperturbative parts. In general, we expect that we can choose
either dirac structure $\Pi_{i}$(with i =1,2,3,4) of the
correlations $\Pi(p,p',q)$ to study the hadronic coupling constants.
In our calculations, we observe that the structure
$p\!\!/\gamma_{5}$ and $q\!\!/\gamma_{5}$ are the pertinent dirac
structures.

After taking its imaginary parts of $\Pi_{i}$, we get the spectral
densities $\rho_{i}(s,s',Q^2)$ of the corresponding Dirac structure.
Using dispersion relations, each $\Pi_{i}$ can be written as:
\begin{eqnarray}
\Pi^{OPE}_{i}(Q^{2})=\int ds\int
ds'\frac{\rho_{i}^{pert}(s,s',Q^{2})+\rho_{i}^{non-pert}(s,s',Q^{2})}{(s-p^{2})(s'-p'^{2})}
\end{eqnarray}
Where $s=p^2$, $s'=p'^2$ and $Q^2=-q^2$. As examples, we give the
perturbative and nonperturbative parts of the spectral densities for
the two Dirac structures $p\!\!\!/\gamma_{5}$ and
$q\!\!\!/\gamma_{5}$
\begin{eqnarray}
\notag\
\rho_{p\!\!\!/\gamma_{5}}^{pert}(s,s',Q^2)&&=\int^{1}_{0}dx\int^{1-x}_{0}dy\frac{-1}{32\pi^{4}(x+y-1)}\Big\{-
\Big[2 m_{b[c]} (x + y) + m_{d} (3 x + 3 y - 1) -m_{u} (x + y - 1)\Big] \\
\notag\ && \times\Big[x (m_{b[c]}^2 + Q^2y) + sx (x + y - 1)  +
s'y(x + y - 1)\Big] + s (x + y)\Big [m_{b[c]} x
(2 x + 2 y - 1) \\
\notag\ && + 3 m_{d} (x - 1) (x + y - 1)  -m_{u}(x^2 +
x(y-4)-2y+3)\Big] +s'(x +y) \Big[m_{b[c]} \Big(x (2 y - 1) \\
\notag\ &&+ 2 (y - 1) y\Big) +y \Big(3 m_{d} (x + y - 1)  - m_{u} (x
+ y - 2)\Big)\Big] + 9 m_{b[c]} m_{d} m_{u} x +9 m_{b[c]} m_{d}
m_{u} y \\
\notag\ &&- 6 m_{b[c]} m_{d} m_{u} - 3 m_{b[c]} m_{u}^2 x  - 3
m_{b[c]} m_{u}^2 y  + 2 m_{b[c]} Q^2 x^2 y - m_{b[c]} Q^2 x^2 + 2
m_{b[c]} Q^2 x y^2 \\
\notag\ && - m_{b[c]} Q^2 x y - 6 m_{d} m_{u}^2 x - 6 m_{d} m_{u}^2
y + 6 m_{d} m_{u}^2  + 3 m_{d} Q^2 x^2 y + 3 m_{d} Q^2 x y^2 - 3
m_{d} Q^2 x y \\
\notag\ && - 3 m_{d} Q^2 y^2 + 3 m_{d} Q^2 y - m_{u} Q^2 x^2 y -
m_{u} Q^2 x y^2 + 2 m_{u} Q^2 x y + 2 m_{u} Q^2 y^2 - 3 m_{u} Q^2
y\Big\}\\
\notag\ && \times\Theta[H_{2}(s,s',Q^{2})]
\end{eqnarray}
\begin{eqnarray}
\notag\
\rho_{q\!\!\!/\gamma_{5}}^{pert}(s,s',Q^2)&&=\int^{1}_{0}dx\int^{1-x}_{0}dy\frac{1}{32\pi^{4}(x+y-1)^{2}}
\Big\{(x + y - 1) \Big[s \Big(m_{b[c]} x y (2 x + 2 y - 1) + 3 m_{d} (x^2 (y - 1) \\
\notag\ && +x(y^2-3y + 1) - (y - 1) y) - m_{u} (x^2 (y - 3) + x (y^2
-
7 y + 3) + (3 - 2 y) y)\Big) \\
\notag\ &&  + s' y \Big(m_{b[c]} (x (2 y - 1) + 2 (y - 1) y) + 3
m_{d} (y - 1) (x
+ y - 1) - m_{u} (x (y - 3) + y^2 - 5 y + 3)\Big) \\
\notag\ &&  + 9 m_{b[c]} m_{d} m_{u} y - 6 m_{b[c]} m_{d} m_{u} - 3
m_{b[c]} m_{u}^2 y + 2 m_{b[c]} Q^2 x
y^2 - m_{b[c]} Q^2 x y - 6 m_{d} m_{u}^2 y   + 3 m_{d} Q^2 x y^2 \\
\notag\ && - 3 m_{d} Q^2 x y - 3 m_{d} Q^2 y^2 + 3 m_{d} Q^2 y -
m_{u} Q^2 x y^2 + 3
m_{u} Q^2 x y + 2 m_{u} Q^2 y^2 - 3 m_{u} Q^2 y\Big] \\
\notag\ && - \Big[m_{b[c]} (x (6 y - 1) + 6 (y - 1) y) + 3 m_{d} (3
y - 2) (x + y
- 1) + m_{u} (-3 x (y - 2) - 3 y^2 + 10 y - 6)\Big] \\
\notag\ && \times\Big[x (m_{b[c]}^2 + Q^2 y) + s x (x + y - 1) + s'
y (x + y - 1)\Big]\Big\}\Theta[H_{2}(s,s',Q^{2})]
\end{eqnarray}
\begin{eqnarray}
\notag\
\rho_{p\!\!\!/\gamma_{5}}^{non-pert}(s,s',Q^2)&&=\int^{1}_{0}dx\int^{1-x}_{0}dy\frac{3(\langle
u\overline{u}\rangle-\langle d\overline{d} \rangle)}{12\pi^{2}}
\Big(3 x
+ 3 y - 1\Big)\Theta[H_{2}(s,s',Q^{2})]\\
\notag\ && -\frac{\langle
u\overline{u}\rangle}{48\pi^{2}(m_{b[c]}^{2}+Q^{2})^2}\Big\{6
m_{b[c]}^3 m_{d} - 12
m_{b[c]}^3 m_{u}  - 6 m_{b[c]}^2 m_{d} m_{u} + 6 m_{b[c]}^2 m_{u}^2 \\
\notag\ && + s' \Big[2 m_{b[c]}^2 - m_{b[c]} m_{u} + 2 Q^2\Big]  - 2
m_{b[c]}^2 Q^2 + 6 m_{b[c]} m_{d} Q^2
+ s \Big[m_{b[c]} (m_{u} - 2 m_{b[c]}) - 2 Q^2\Big] \\
\notag\ && - 11 m_{b[c]} m_{u} Q^2 - 3 m_{d} m_{u} Q^2 + 3 m_{u}^2
Q^2 - 2
Q^4\Big\}\Theta[ H_{1}[s,s',Q^2]]\\
\notag\ && -\int^{1}_{0}dx\int^{1-x}_{0}dy\langle
\alpha_{s}\frac{G^{2}}{\pi}\rangle\frac{x^{3}(3x+3y-2)m_{b[c]}}{16\pi^{2}(x+y-1)}\delta
[H_{2}[s,s',Q^2]]
\\
\notag\ && +\frac{\langle qqg
\rangle}{8\times4\pi^{2}\times9(m_{b[c]}^{2}+Q^{2})^4}\Big\{18
m_{b[c]}^6 - 18
m_{b[c]}^5 m_{d} - 36 m_{b[c]}^5 m_{u} - 18 m_{b[c]}^4 m_{d} m_{u} + 18 m_{b[c]}^4 m_{u}^2 \\
\notag\ && +
   39 m_{b[c]}^4 Q^2 - 36 m_{b[c]}^3 m_{d} Q^2 - 32 m_{b[c]}^3 m_{u} Q^2 + 30 m_{b[c]}^2 Q^4 -
   3 s \Big[m_{b[c]}^3 (3 m_{b[c]} - 2 m_{u}) \\
\notag\ && + 4 m_{b[c]}^2 Q^2 + Q^4\Big] +
   3 s' \Big[3 m_{b[c]}^4 - 2 m_{b[c]}^3 m_{u}  + 4 m_{b[c]}^2 Q^2 + Q^4\Big] -
   18 m_{b[c]} m_{d} Q^4 \\
\notag\ && - 2 m_{b[c]} m_{u} Q^4 + 9 Q^6\Big\}\Theta[
H_{1}[s,s',Q^2]]
\end{eqnarray}
\begin{eqnarray}
\notag\
\rho_{q\!\!\!/\gamma_{5}}^{non-pert}(s,s',Q^2)&&=\int^{1}_{0}dx\int^{1-x}_{0}dy\frac{3}{16\times12\pi^{2}}
\Big\{(48 y - 32)\langle d\overline{d} \rangle+(32 - 48 y)\langle
u\overline{u})\Big\}\Theta[H_{2}(s,s',Q^{2})]\\
\notag\ && -\frac{\langle
u\overline{u}\rangle}{48\pi^{2}(m_{b[c]}^{2}+Q^{2})^2}\Big\{-3
m_{b[c]} \Big[2 m_{b[c]}^2 (m_{d} -
2 m_{u}) + m_{b[c]} m_{u} (m_{u} - 5 m_{d}) + 2 m_{d} m_{u}^2\Big] \\
\notag\ && +
  Q^2 \Big[2 m_{b[c]}^2 - 6 m_{b[c]} m_{d} + 11 m_{b[c]} m_{u} +
     12 m_{d} m_{u}\Big] + s \Big[m_{b[c]} (2 m_{b[c]} - m_{u}) + 2 Q^2\Big] - 2 Q^4\Big\}\\
\notag\ && \times\Theta[ H_{1}[s,s',Q^2]]
      +\int^{1}_{0}dx\int^{1-x}_{0}dy\langle
\alpha_{s}\frac{G^{2}}{\pi}\rangle\frac{x^{3}(18 x y-x+18 y^2-18
y+2)m_{b[c]}}{96\pi^{2}(x+y-1)^{2}}\delta
[H_{2}[s,s',Q^2]] \\
\notag\ && -\frac{\langle qqg
\rangle}{8\times4\pi^{2}\times9(m_{b[c]}^{2}+Q^{2})^4} \Big\{27
m_{b[c]}^6 - 54
m_{b[c]}^5 m_{d} - 54 m_{b[c]}^5 m_{u} - 72 m_{b[c]}^4 m_{d} m_{u} + 18 m_{b[c]}^4 m_{u}^2 \\
\notag\ && -
   60 m_{b[c]}^4 Q^2 +
   36 m_{b[c]}^3 m_{d} m_{u}^2 + s \Big[m_{b[c]}^3 (-15 m_{b[c]} + 36 m_{d} + 10 m_{u}) \\
\notag\ &&-
      4 m_{b[c]} Q^2 (6 m_{b[c]} - 9 m_{d} - m_{u})  - 9 Q^4\Big] +
   2 s' \Big[m_{b[c]}^3 (3 m_{b[c]} - 72 m_{d} - 2 m_{u}) \\
\notag\ &&-
      2 m_{b[c]} Q^2 (-3 m_{b[c]} + 9 m_{d} + m_{u}) + 3 Q^4\Big]  - 72 m_{b[c]}^3 m_{d} Q^2 -
   64 m_{b[c]}^3 m_{u} Q^2 \\
\notag\ &&- 72 m_{b[c]}^2 m_{d} m_{u} Q^2 + 45 m_{b[c]}^2 Q^4 -
   18 m_{b[c]} m_{d} Q^4 - 16 m_{b[c]} m_{u} Q^4 - 18 m_{d} m_{u} Q^4 \\
\notag\ &&+ 12 Q^6\Big\}\Theta[ H_{1}[s,s',Q^2]]
\end{eqnarray}
Where $\Theta$ denotes the unit-step function, and
$H_{1}[s,s',Q^2]$, $H_{2}[s,s',Q^2]$ are defined as:
\begin{eqnarray}
&& H_{1}[s,s',Q^2]=s' \\
&&
 H_{2}[s,s',Q^2]=x (m_{b[c]}^2+Q^2 y)+s x (x+y-1)+s' y (x+y-1)
\end{eqnarray}

\begin{large}
\textbf{2.3 The strong coupling constant}
\end{large}

We perform a double Borel transformation\cite{Cola} to the
physiological as well as the OPE sides. Then, we equate these two
sides, invoking the quark-hadron duality from which the sum rule is
obtained. As an example, the form factors for the structure
$p\!\!/\gamma_{5}$ is:
\begin{eqnarray}
&&G^{p\!\!/\gamma_{5}}_{\Sigma_{b}NB[\Sigma_{c}ND]}(Q^{2})=e^{\frac{m_{\Sigma_{b}[\Sigma_{c}]}^{2}}{M1^{2}}}e^{\frac{m_{N}^{2}}{M2^{2}}}\frac{(m_{b[c]}+m_{u})(Q^{2}+m^{2}_{B[D]})}{m_{B[D]}^{2}f_{B[D]}\lambda_{\Sigma_{b}[\Sigma_{c}]}\lambda_{N}(m_{\Sigma_{b}[\Sigma_{c}]}-m_{N})}\\
&& \notag \times\Big\{ \int^{s_{0}}_{(m_{b[c]}+m_{u}+m_{d})^{2}} ds
\int^{u_{0}}_{(2m_{u}+m_{d})^{2}}
ds'e^{-\frac{s}{M1^{2}}}e^{-\frac{s'}{M1^{2}}}\Big[\rho_{p\!\!/\gamma_{5}}^{pert}(s,s',Q^{2})+\rho_{p\!\!/\gamma_{5}}^{non-pert}(s,s',Q^2)\Big]\Big\}
\end{eqnarray}
Where $M_{1}$ and $M_{2}$ are the Borel parameters, $s_{0}$ and
$u_{0}$ are two continuum threshold parameters which are introduced
to eliminate the $h.r.$ terms. These parameters fulfill the
following relations:$m_{i}^{2}$$<$$s_{0}$$<$$m'^{2}_{i}$ and
$m_{o}^{2}$$<$$u_{0}$$<$$m'^{2}_{o}$, where $m_{i}$ and $m_{o}$ are
the masses of the incoming and out-coming hadrons respectively and
$m'$ is the mass of the first excited state of these hadrons.

\begin{large}
\textbf{3 The results and descussions}
\end{large}

Present section is devoted to the numerical analysis of the sum
rules for the coupling constants. The decay constants parameters
used in this work are taken as
$f_{B}=$($248\pm23_{exp}\pm25_{Vub}$)MeV\cite{Khod4},
$f_{D}=$$(205.8\pm8.5\pm2.5$)MeV\cite{Eise},
$\lambda_{N}=$$(0.0011\pm0.0005$)$GeV^{6}$\cite{Azi3},
$\lambda_{\Sigma_{b}}=$$(0.062\pm0.018)$$GeV^{3}$\cite{Azi4} and
$\lambda_{\Sigma_{c}}=$$(0.045\pm0.015)$$GeV^{3}$\cite{Azi4}. We
take the masses of the hadronic from reference\cite{Oliv}, where
$m_{B}=$$(5279.26\pm0.17)$MeV, $m_{D}=$$(1864.84\pm0.07)$MeV,
$m_{N}=$$(938.272046\pm0.000021)$MeV,
$m_{\Sigma_{b}}=$$(5811.3\pm1.9)$MeV,
$m_{\Sigma_{c}}=$$(2452.9\pm0.4)$MeV and of quark
$m_{b}=$($4.18\pm0.03$)GeV, $m_{c}=$($1.275\pm0.025$)GeV,
$m_{d}=$($4.8^{+0.5}_{-0.3}$)MeV, $m_{u}=$($2.3^{+0.7}_{-0.5}$)MeV.
The vacuum condensates are taken to be the standard values
$\langle\overline{u}u\rangle=\langle\overline{d}d\rangle=-(0.8\pm0.1)\times(0.24\pm0.01GeV)^3$\cite{Ioff},
$\langle\overline{s}g_{s}\sigma
Gs\rangle=m_{0}^{2}\langle\overline{s}s\rangle$\cite{Ioff},
$m_{0}^{2}=(0.8\pm0.1)GeV^2$, $\langle
g_{s}^{2}GG\rangle=(0.022\pm0.004)GeV^4$\cite{Bely}. From Eq.(20),
we also know that the value of the form factor
$G_{\Sigma_{b}NB[\Sigma_{c}ND]}$ is the function of the input
parameters, including the Borel parameters $M_{1}^{2}$ and
$M_{2}^{2}$,  continuum threshold $s_{0}$ and $u_{0}$, the momentum
$Q^2$.

The working regions for the $M_{1}^{2}$ and $M_{2}^{2}$ are
determined by requiring not only that the contributions of the
higher states and continuum be effectively suppressed, but also that
the contributions of the higher-dimensional operators are small. In
other words, we should find a good plateau which will ensure OPE
convergence and the stability of our results\cite{Cola}. The plateau
is often called "Borel window". Considering these factors, the Borel
windows are chosen as $7(3)GeV^{2}\leq M_{1}^{2}<14(7)GeV^{2}$ and
$3(2)GeV^{2}\leq M_{2}^{2}<7(6)GeV^{2}$ for the strong vertex
$\Sigma_{b}NB(\Sigma_{c}ND)$(see Figures1-4). From these figures, we
can see that the values are rather stable with variations of the
Borel parameters, it is reliable to extract the form factors. In
addition, the continuum parameters,
$s_{0}=(m_{i}+\bigtriangleup_{i})^{2}$ and
$u_{0}=(m_{o}+\bigtriangleup_{o})^{2}$ are employed to include the
pole and to suppress the $h.r.$ contributions. The values for
$\triangle_{i}$ and $\triangle_{o}$ can not be far from the
experimental value of the distance between the pole and the first
excited state\cite{Cola}. In general, these two continuum thresholds
$s_{0}$ and $u_{0}$ are determined by the relations
$s_{0}\sim(m_{i}+0.5GeV)^{2}$ and $u_{0}\sim(m_{o}+0.5GeV)^{2}$.
According to these considerations, we take $s_{0}=37.4(7.6)GeV^{2}$
and $u_{0}=1.99(1.99)GeV^{2}$ for the strong vertex
$\Sigma_{b}NB(\Sigma_{c}ND)$.

\begin{figure}[h]
\begin{minipage}[t]{0.45\linewidth}
\centering
\includegraphics[height=5cm,width=7cm]{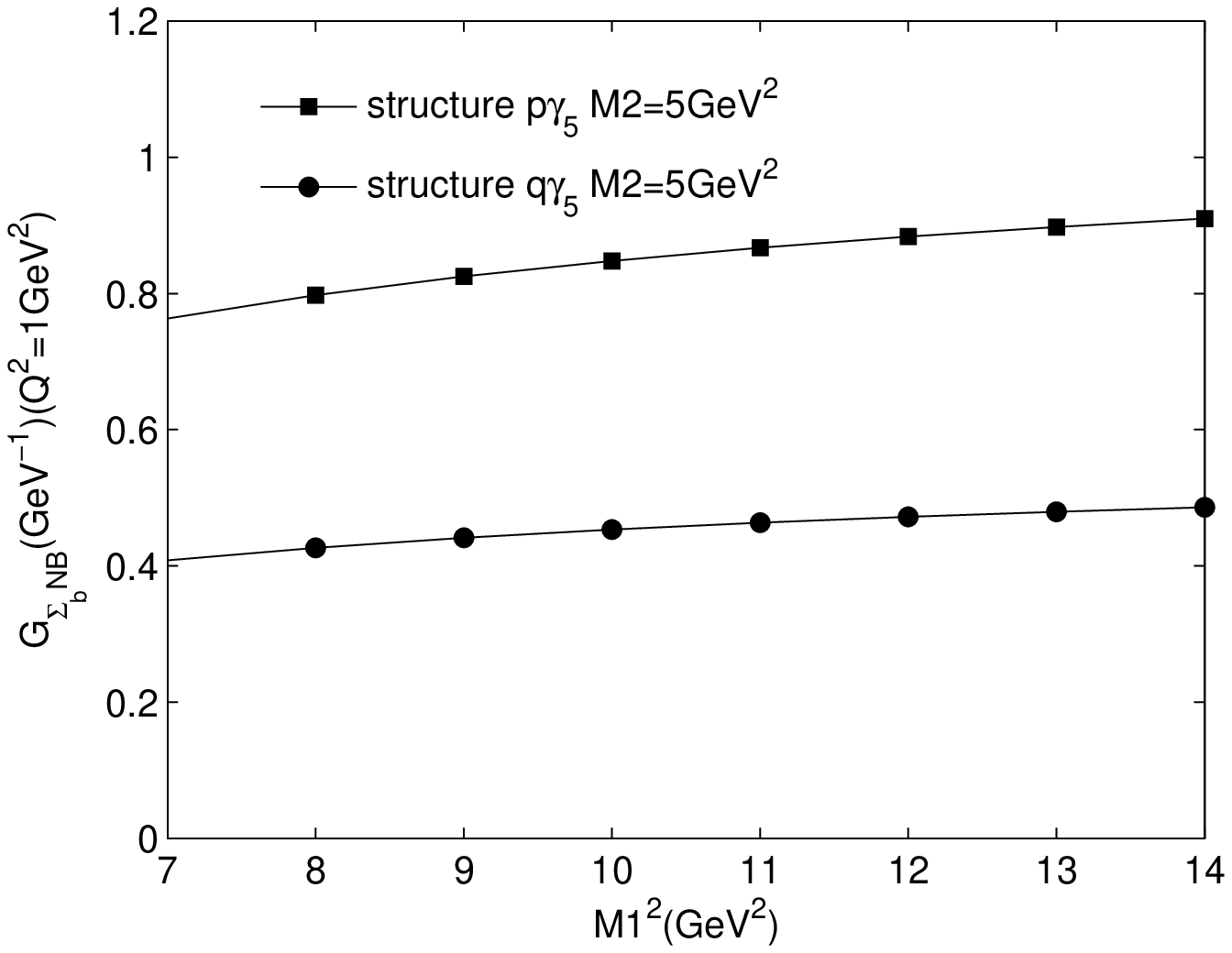}
\caption{$G_{\Sigma_{b} NB}$ as a function of $M_1^2$ at average
values of the continuum thresholds.\label{your label}}
\end{minipage}
\hfill
\begin{minipage}[t]{0.45\linewidth}
\centering
\includegraphics[height=5cm,width=7cm]{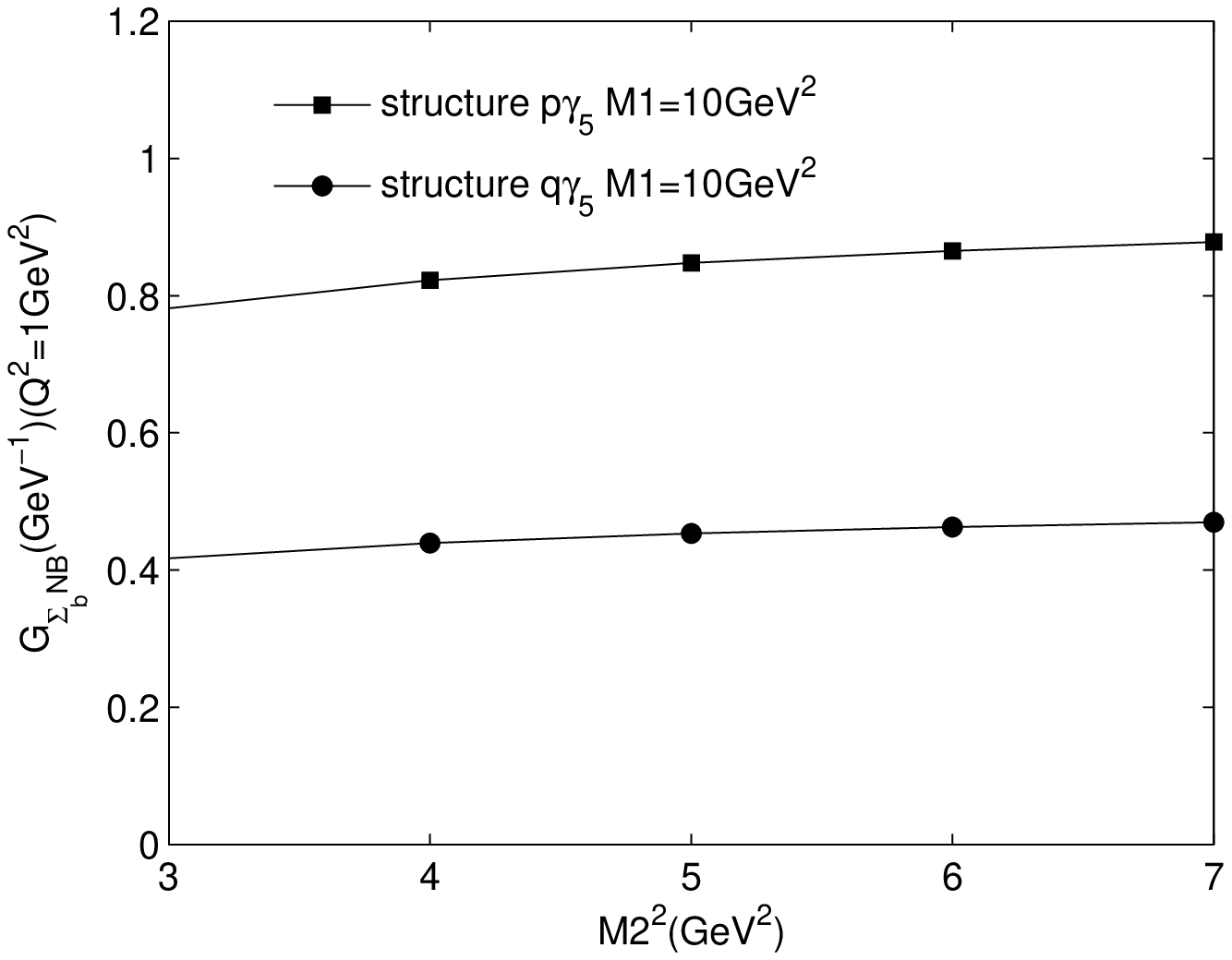}
\caption{$G_{\Sigma_{b} NB}$ as a function of $M_2^2$ at average
values of the continuum thresholds.\label{your label}}
\end{minipage}
\end{figure}
\begin{figure}[h]
\begin{minipage}[t]{0.45\linewidth}
\centering
\includegraphics[height=5cm,width=7cm]{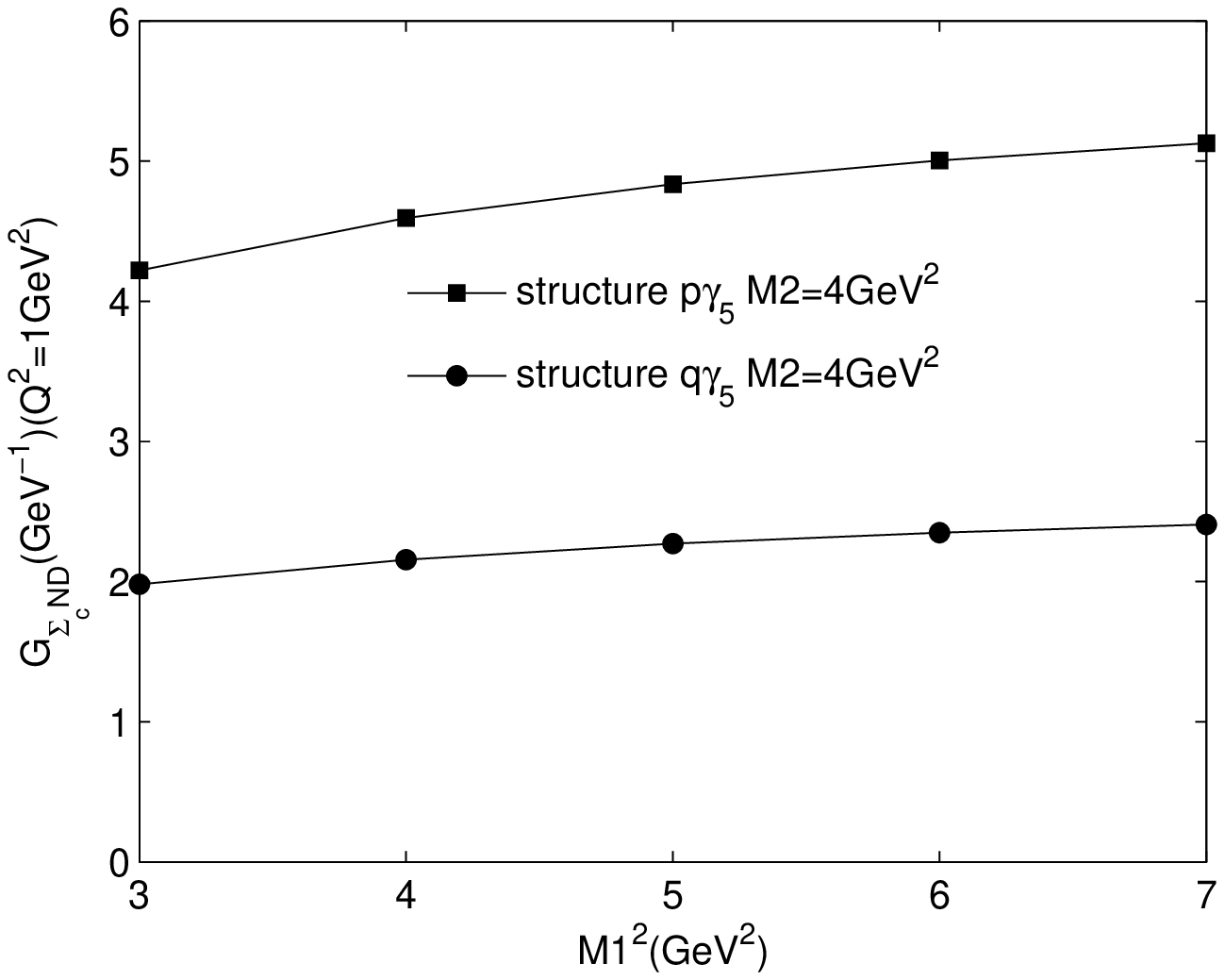}
\caption{$G_{\Sigma_{c} ND}$ as a function of $M_1^2$ at average
values of the continuum thresholds.\label{your label}}
\end{minipage}
\hfill
\begin{minipage}[t]{0.45\linewidth}
\centering
\includegraphics[height=5cm,width=7cm]{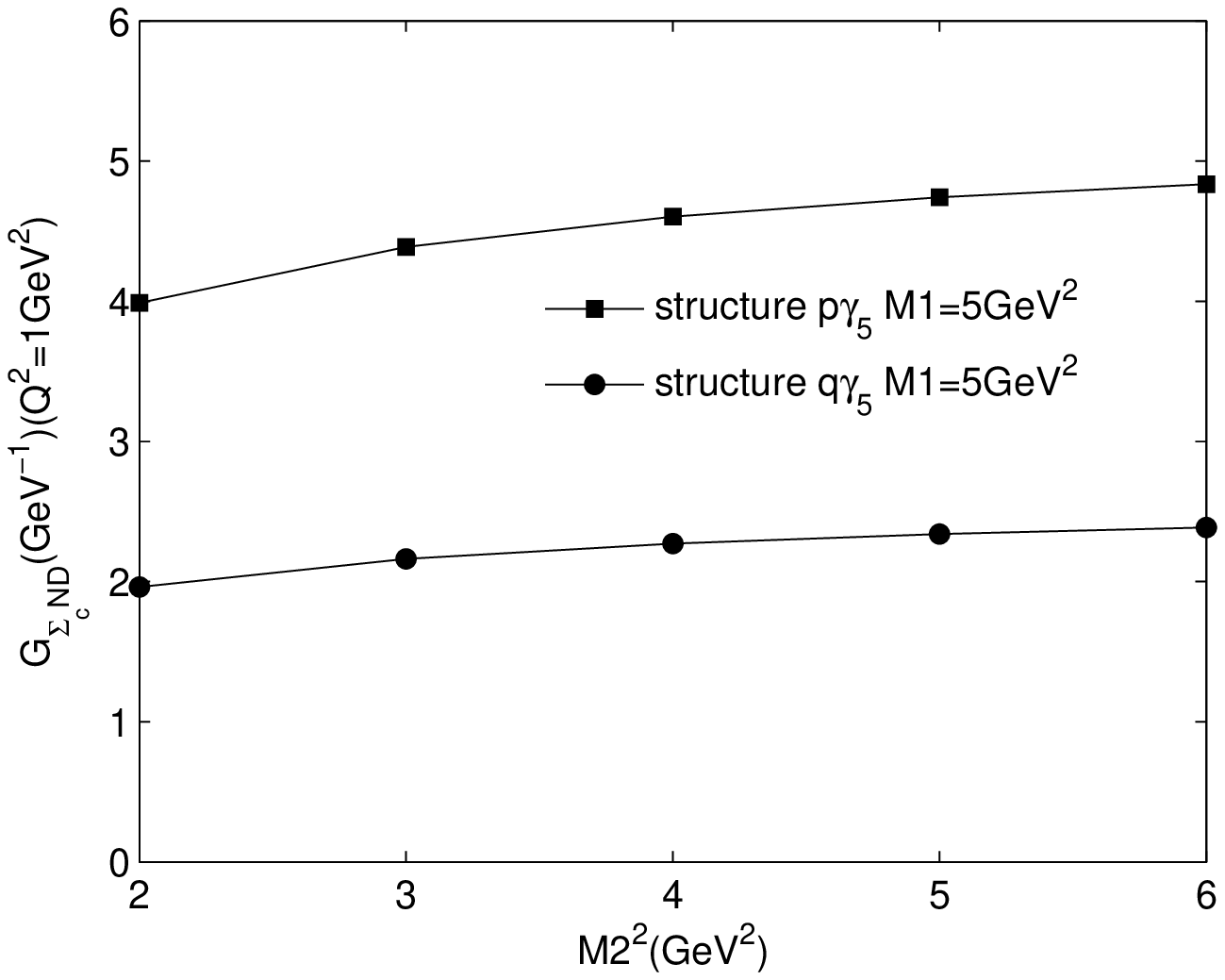}
\caption{$G_{\Sigma_{c} ND}$ as a function of $M_2^2$ at average
values of the continuum thresholds.\label{your label}}
\end{minipage}
\end{figure}

\begin{figure}[h]
\begin{minipage}[t]{0.45\linewidth}
\centering
\includegraphics[height=5cm,width=7cm]{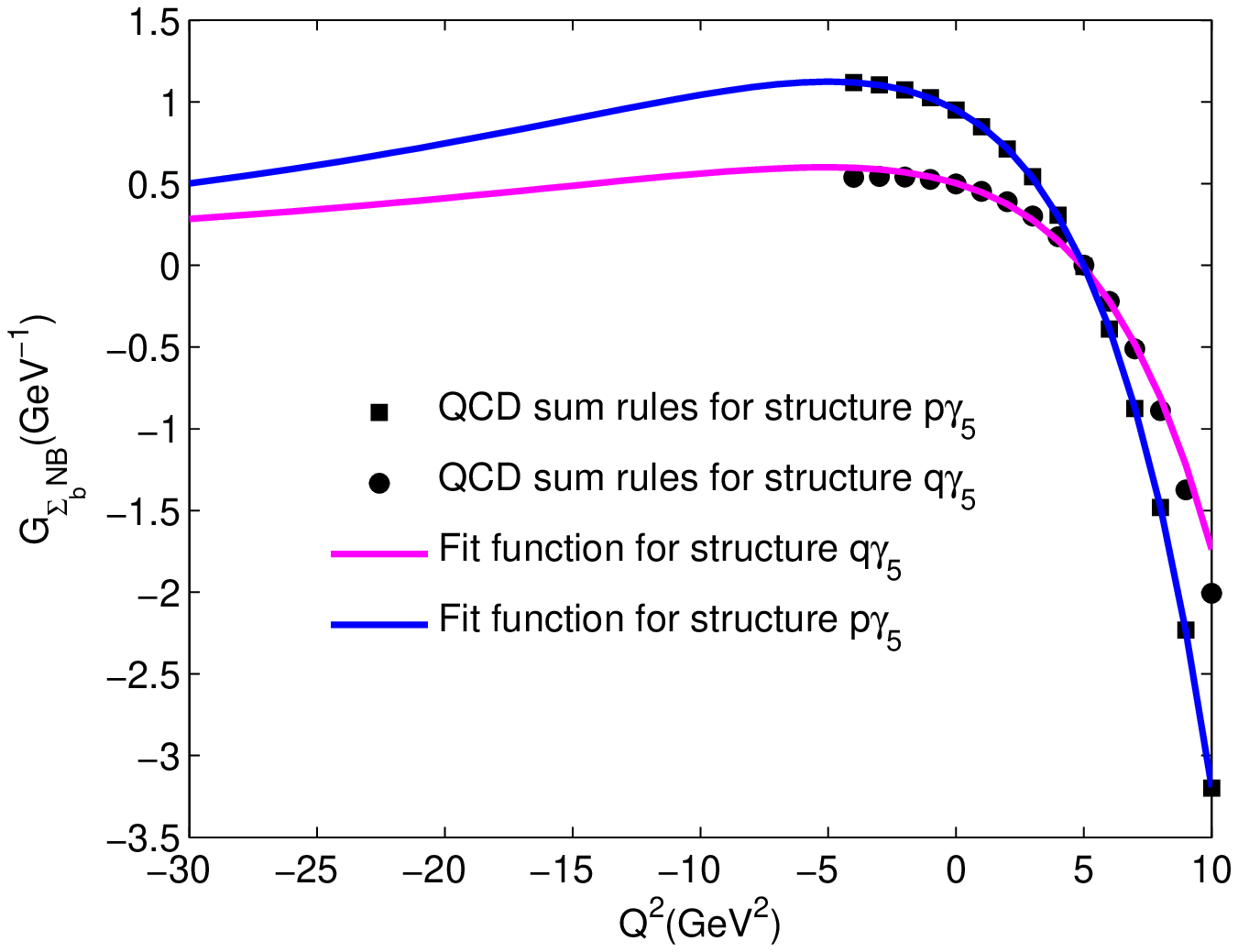}
\caption{$G_{\Sigma_{b} NB}$ as a function of $Q^2$ at average
values of the continuum thresholds and Borel mass
parameters.\label{your label}}
\end{minipage}
\hfill
\begin{minipage}[t]{0.45\linewidth}
\centering
\includegraphics[height=5cm,width=7cm]{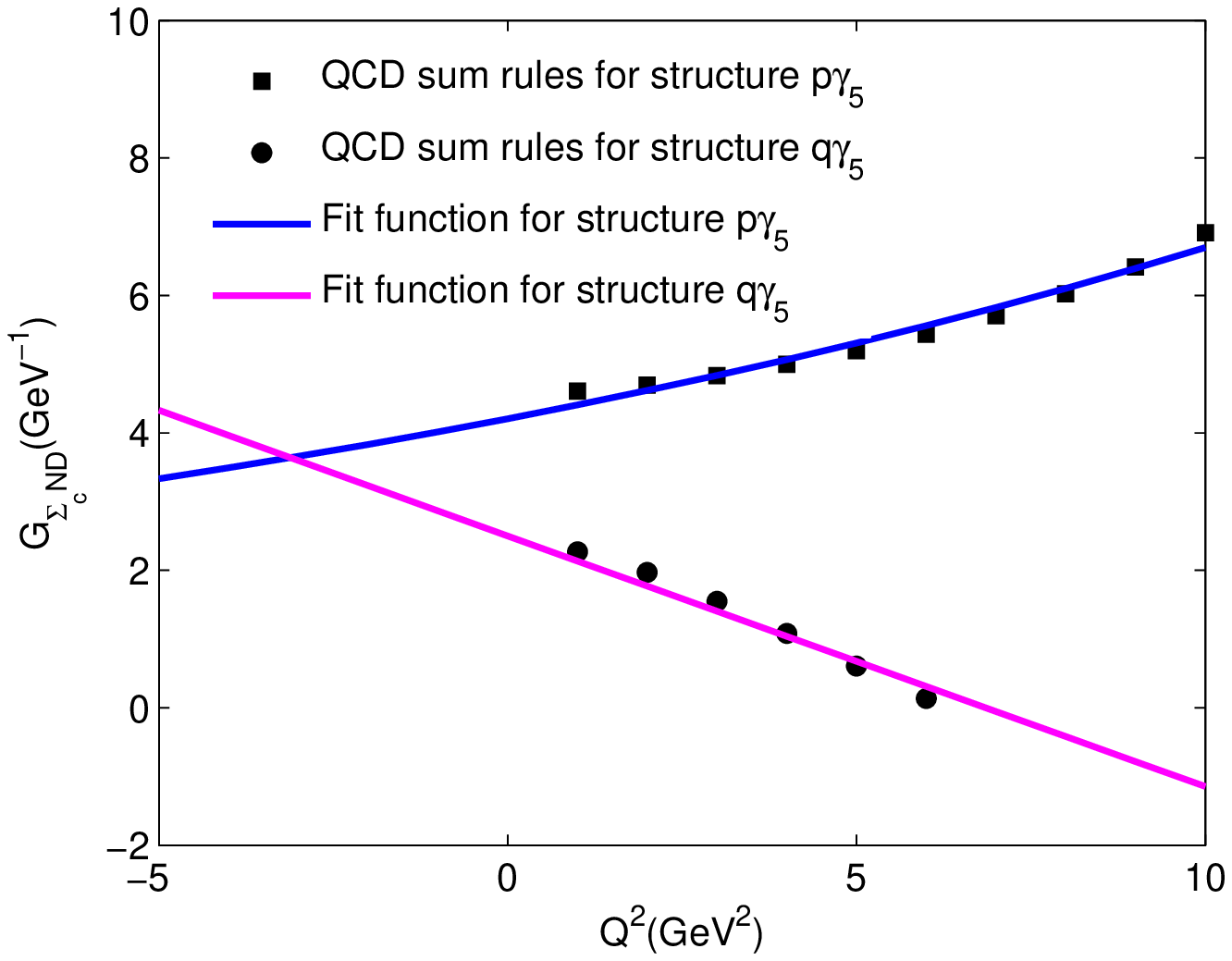}
\caption{$G_{\Sigma_{c} ND}$ as a function of $Q^2$ at average
values of the continuum thresholds and Borel mass
parameters.\label{your label}}
\end{minipage}
\end{figure}
However, in order to obtain the coupling constants, it is necessary
to extrapolate these results into physical regions($Q^{2}<0$), which
is realized by fit the form factors into suitable analytical
functions. It is indicated that we should get the same values for
the coupling constants for the different dirac structure
$p\!\!/\gamma_{5}$ or $q\!\!/\gamma_{5}$ when we take
$Q^{2}=-m_{B[D]}^{2}$. This above procedure can help us minimizing
the uncertainties in the calculation of the coupling constant, which
will be quite clear in the following section. From our analysis, we
observe that the dependence of the form factors on $Q^{2}$ can be
well described by the following fit function(see Figures5-6 ):
\begin{eqnarray}
G_{\Sigma_{b}NB[\Sigma_{c}ND]}(Q^{2})=C_{1}exp^{-\frac{Q^2}{C_{2}}}+C_{3}exp^{-\frac{Q^2}{C_{4}}}
\end{eqnarray}

Where the values of $C_{1}$,$C_{2}$,$C_{3}$ and $C_{4}$ for
different dirac structures are presented in Table 1 for these two
strong vertexes $\Sigma_{b}NB$ and $\Sigma_{c}ND$. The fit function
is used to determine the value of the strong coupling constant at
$Q^{2}=-m_{B[D]}^{2}$ for different structures, and the results are
also presented in Table 1. The errors existing in these results
arise from the uncertainties of the input parameters together with
the uncertainties coming from the determination of the working
regions of the auxiliary parameters.


It is indicated from Figure 5, Figure 6 and Table 1 that different
dirac structure can give compatible results for each strong coupling
constants when we take $Q^{2}=-m_{B[D]}^{2}$ in the fit
function(Eq.21). For example, the results of the coupling constant
for vertex $\Sigma_{b}NB$ are $0.55$ and $0.31$ for
$p\!\!/\gamma_{5}$ and $q\!\!/\gamma_{5}$ structure respectively.
Thus, we can take the average of the coupling constants from two
different dirac structure for each vertex, which are $0.43\pm0.01$
and $3.76\pm0.05$ for $\Sigma_{b}NB$ and $\Sigma_{c}ND$
respectively.

\begin{table*}[t]
\begin{ruledtabular}\caption{Parameters appearing in the fit function of the coupling form factor for $\Sigma_{b}NB$ and $\Sigma_{c}ND$.}
\begin{tabular}{c c c c c c c c c c c c c c c c c c}
& \ Structure  & \ $C_{1}$($GeV^{-1}$)  & \ $C_{2}$($GeV^{2}$)  & \ $C_{3}$($GeV^{-1}$) & \ $C_{4}$ ($GeV^{2}$) & \ $G$ \\
\hline
 $\Sigma_{b}NB$ &  \ $p\!\!/\gamma_{5}$       &  \  $-0.40\pm0.05$      & \  $-4.89\pm0.02$       &  \   $0.90\pm0.05$   & \  $-26.13\pm4.00$  & \ $0.55 \pm0.01$    \\
&  \ $q\!\!/\gamma_{5}$       &  \  $-0.78\pm0.10$      & \  $-4.99\pm0.02$       &  \   $1.74\pm0.10$   & \  $-24.22\pm4.00$   & \ $0.31\pm0.01$   \\
\hline
$\Sigma_{c}ND$&  \ $p\!\!/\gamma_{5}$       &  \  $4.21\pm0.20$      & \  $-21.51\pm3.00$       &  \   $0$   & \  $0$   & \ $3.58\pm0.02$   \\
&  \ $q\!\!/\gamma_{5}$       &  \  $663.20\pm63.50$      & \  $1728.30\pm25.60$       &  \   $-660.60\pm68.70$   & \  $0$   & \ $3.94\pm0.04$   \\
\end{tabular}
\end{ruledtabular}
\end{table*}

\begin{large}
\textbf{4 Conclusion}
\end{large}

In this article, we have calculated the form factors of the vertexes
$\Sigma_{b}NB$ and $\Sigma_{c}ND$ in the space-like regions by
three-point sum rules. Then we fit the form factors into analytical
functions, extrapolated them into the time-like regions, and
obtained the strong coupling constants $G_{\Sigma_{b}NB}$ and
$G_{\Sigma_{c}ND}$. These calculated results can be used to analyze
the related experimental results at LHC as well as the heavy ion
collision experiments like $\overline{P}ANDA$ at FAIR.


\begin{large}
\textbf{Acknowledgment}
\end{large}

This work has been supported by the Fundamental Research Funds for
the Central Universities, Grant Number $2016MS133$.

\end{document}